\documentclass[aps,prb,twocolumn,groupedaddress,amsmath,amssymb]{revtex4}

\usepackage{amssymb}
\usepackage{amsmath}
\usepackage{graphicx}
\usepackage{subfigure}
\usepackage{textcomp}
\usepackage{color}
\usepackage{amsfonts}
\usepackage{bbold}
\usepackage{dsfont}
\usepackage{epsfig}
\usepackage{hyperref}

\newcommand{\arctanh}[1]{\operatorname{arctan}}

\begin{document}

\title{Bias asymmetry in the conductance profile of magnetic ions on surfaces probed by scanning tunneling microscopy}

\author{Aaron Hurley, Nadjib Baadji and Stefano Sanvito}
\affiliation{School of Physics and CRANN, Trinity College, Dublin 2, Ireland}

\date{\today}

\begin{abstract}
The conductance profiles of magnetic transition metal atoms, such as Fe, Co and Mn, deposited on surfaces and probed by a scanning tunneling 
microscope (STM), provide detailed information on the magnetic excitations of such nano-magnets. In general the profiles are symmetric
with respect to the applied bias. However a set of recent experiments has shown evidence for inherent asymmetries when either a normal or a 
spin-polarized STM tip is used. In order to explain such asymmetries here we expand our previously developed perturbative approach to 
electron-spin scattering to the spin-polarized case and to the inclusion of out of equilibrium spin populations. In the case of a magnetic STM tip 
we demonstrate that the asymmetries are driven by the non-equilibrium occupation of the various atomic spin-levels, an effect that reminds closely 
that electron spin-transfer. In contrast when the tip is not spin-polarized such non-equilibrium population cannot be build up. In this circumstance 
we propose that the asymmetry simply originates from the transition metal ion density of state, which is included here as a non-vanishing real 
component to the spin-scattering self-energy. 
\end{abstract}


\maketitle

\section{Introduction}
The possibility of altering and controlling the spin-state of a single magnetic ion or of a small magnetic cluster with an external probe
represents a unique opportunity towards the understanding and the exploitation of the magnetic interaction at the most microscopic level. 
Possible areas of application for such ability may include spin-based quantum logic, where one necessitates to prepare, manipulate and 
read spin-qubits. It is then crucial to develop tools capable of addressing the single spin-limit. Low-temperature scanning tunneling 
microscopy provides one of such tools. In general the method exploits a scanning tunneling microscope (STM) operated in spectroscopical 
mode, by which the inelastic electron tunneling spectroscopy (IETS) at the spin-excitations of a given system is measured \cite{Heinrich2004}. 
This scheme is known as spin-flip IETS (SF-IETS). The same STM can also be used to position and manipulate the magnetic atoms on a 
non-magnetic substrate \cite{Eigler}, so that STM appears both as a fabrication and subtle characterization tool.

Transition metal magnetic atoms on insulating surfaces, in particular Mn\cite{Hir1,Loth1}, Co\cite{Otte1,Parks} and Fe\cite{Hir2,Otte2,Loth2}, have been 
the focus of intensive research in the last few years. These have all partially filled $d$-shells, which are highly localized and responsible for the 
magnetic moment, and extended $s$-like electrons, which are responsible for the electron conduction. In general $s$ and $d$ electrons 
interact via exchange coupling so that the magnetic structure is coupled to the conducting electrons. The magnetic atoms are usually deposited 
on carefully prepared CuN-decorated Cu surfaces, where the typical electronic coupling is weak enough that the magnetism is preserved, but 
it is sufficiently strong to break the atomic central symmetry so that magnetic anisotropy develops. STM experiments are then conducted and 
the fingerprint of a magnetic excitation is a step in the differential conductance, $G=\mathrm{d}I/\mathrm{d}V$, as a function of bias, $V$ ($I$ is
the STM current). These appear at the critical voltage necessary to open a new inelastic transport channel, i.e. at voltages corresponding to the
given magnetic excitation energy. 

Several methods aimed at modeling SF-IETS have been recently developed. Early theoretical work has focused on second order 
perturbation theory to describe the experimental conductance spectra of equilibrium spins by either using a master equation approach 
\cite{Chen, Romeike, HT, Fernandez-Rossier, Lorente, Persson} or a non-equilibrium Green's function one \cite{Hurley}. More recently this scheme has 
been extended to third order, which allows us to describe additional features in the $G(V)$ line-shape that cannot be accounted for at the 
second order level \cite{HurleyKondo, Elste, Zitko1, Zitko2}. These works have been very successful in describing conductance profiles, 
which appear symmetric with the external bias polarity, i.e. that they cannot distinguish whether the current flows from the sample to the tip 
or in the opposite direction. However, recent experiments have shown that regardless on whether a non-spin-polarized \cite{Hir1} or a 
spin-polarized \cite{Loth1,Loth2} tip is used the IETS profiles exhibit an intrinsic asymmetry with respect to the applied voltage, i.e. $G(V)\ne G(-V)$. 

In the case of a spin-polarized STM tip, where the tip density of states is spin split between majority (spin up) and minority (spin down) carriers, 
the asymmetry has been theoretically well explained \cite{Loth1,Loth2,Fransson2,Delgado1,Delgado2,Novaes}. It has been shown that spin selection 
rules enforce a suppression of the inelastic scattering, which depends on the direction of the electrons flow. This results in a asymmetric 
conductance profile, where the magnitude of the asymmetry depends directly on the spin-polarization of the tip. It is also well understood that 
by driving spins out of equilibrium (e.g. by decreasing the tip-sample distance) the conductance line-shape changes 
\cite{Sothmann,Delgado1,Delgado2,Loth1,Loth2}. In this case we must assume that the tunneling electrons influence the spin-state of the atom 
as the time between inelastic events is small compared to the spin relaxation time. A tunneling electron can then encounter the local spin 
in an excited state far from the ground state. The non-equilibrium population of the various accessible spin-states then becomes bias-dependent 
and, for spin-polarized tips, this enhances the asymmetry of the $G(V)$ line-shape. 

Also in the case of a non-spin-polarized tip a bias asymmetry has been revealed experimentally \cite{Hir2}. In particular this appears to 
be quite prominent for both single Mn atoms and Mn mono-atomic chains. This feature has been previously ascribed to a shift in the magnetic 
atom on-site energy, i.e. to an effect arising from the details of the density of states of the atom producing scattering. Such a density of state 
effect produces a non-trivial slope in the conductance as a function of bias \cite{Delgado3}. The on-site energy shift however does not account 
for the asymmetry seen in the inelastic step heights, which also depends on bias. Here we provide an alternative theoretical description, 
which allows us to better fit the experimentally found conductance line-shape.

In our previous works \cite{Hurley,HurleyKondo} we have combined the non-equilibrium Green's function (NEGF) \cite{Keldysh,Datta} 
formalism with a perturbative expansion of the electron-spin interaction in order to describe SF-IETS spectra in a manner, which is fully 
amenable to an implementation within density functional theory (DFT)\cite{Rocha,Rungger}. The scheme essentially consists in constructing 
an electron-spin interaction self-energy, which describes the inelastic tunneling events. The interacting self-energy was previously expanded 
first up to second order~\cite{Hurley} and then to the third order~\cite{HurleyKondo}, with this latter describing the logarithmic decays of $G(V)$ 
at each conductance step. Although both non-equilibrium effects and spin-polarized IETS have been well described up to the second order 
by the master equation approach, in this work we extend our formalism to include the extra line-shape features that can be ascribed to the 
third order self-energy. We also propose that the real part of the interacting self-energy, which has been well studied in the case of 
electron-phonon interactions~\cite{Hyldgaard,Lee}, is a necessary addition to the NEGF formalism in order to account for asymmetric features 
in non-spin polarized systems.

The layout of the paper is as follows. In the next section we extend our NEGF formalism to account for spin polarized leads. In the same 
section we also derive a second-order electron-spin self-energy, which includes a second order expansion of the spin propagator. This provides 
the means to study non-equilibrium effects that result from high current densities. Furthermore, in the case of non-spin polarized tips we 
derive an expression for the real part of the scattering self-energy up to second order. All of the above is combined with the third order 
contribution to the electron propagator. Then we move to the results. First we study the non-equilibrium effects arising from an increase
in current density for the case of a Mn dimer. Then we recreate the spin-polarized experiments of Loth et al. on single Mn and Fe atoms \cite{Loth1,Loth2}. 
Finally we test how the inclusion of the real part of the scattering self-energy modifies the SF-IETS for a non-polarized tip probing Mn 
monomers and trimers.
%


\section{Theoretical Models}
\subsection{Interaction Hamiltonian and contour Green's functions}

We consider here the same single-orbital tight-binding model used in our previous works~\cite{Hurley,HurleyKondo} to describe a magnetic 
system (S) coupled to two non-interacting electrodes representing respectively the STM tip (tip) and the substrate (sub). The scattering 
region containing the magnetic nanostructure consists of a one-dimensional chain of $N$ magnetic atoms. Each of the $i$-th atoms carries 
a quantum mechanical spin $\mathbf{S}_i$ and it is characterized by an on-site energy $\varepsilon_0$. We assume that the tip and the 
substrate can only couple to one atom at a time in the scattering region thus to broaden the electronic level $\varepsilon_0$ through the 
interaction with the electrode by $\Gamma_\mathrm{tip-S/sub-S}$. The Hamiltonian in the scattering region is then described by 
$H_\mathrm{S}={H}_\mathrm{e}+{H}_\mathrm{sp}+{H}_\mathrm{e-sp}$, where ${H}_\mathrm{e}$ is the tight-binding electronic part, 
${H}_\mathrm{sp}$ is the spin part and ${H}_\mathrm{e-sp}$ describes the electron-spin interaction. The various terms can be written
explicitly as
\begin{align}
\label{eq:1}
&{H}_\mathrm{e}=\varepsilon_0\sum_{\lambda\:\alpha}c_{\lambda\alpha}^{\dagger}c_{\lambda\alpha}\:,\\
\label{eq:2}
&{H}_\mathrm{sp}=J_\mathrm{dd}\sum_{\lambda}^{N-1}\mathbf{S}_\lambda\cdot\mathbf{S}_{\lambda+1}+\\ \nonumber&
+\sum_{\lambda}^N\big\{g{\mu_\mathrm{B}}\mathbf{B}\cdot\bold{S}_\lambda+D({S}^z_\lambda)^2+E[({S}^x_\lambda)^2-({S}^y_\lambda)^2]\big\}\:,\\
\label{eq:3}
&{H}_\mathrm{e-sp}=J_\mathrm{sd}\sum_{\lambda\:\alpha,\alpha'}(c_{\lambda\alpha}^{\dagger}[\boldsymbol{\sigma}]_{{\alpha}{\alpha'}}c_{\lambda\alpha'})\cdot\mathbf{S}_\lambda\: \\ \nonumber &
+J_\mathrm{0}\sum_{\lambda\:\alpha}c_{\lambda\alpha}^{\dagger}c_{\lambda\alpha}\:,
\end{align}
where the electron ladder operator $c_{i\alpha}^{\dagger}$ ($c_{i\alpha}$) creates (annihilates) an electron at site $i$ with spin $\alpha$ 
($\alpha=\uparrow,\downarrow$) and on-site energy $\varepsilon_0$.

We model the spin-spin interaction between the localized $\{\mathbf{S}_i\}$ spins by a nearest neighbour Heisenberg Hamiltonian with  
coupling strength $J_\mathrm{dd}$. Furthermore we include interaction with an external magnetic field $\mathbf{B}$ ($\mu_\mathrm{B}$ is
the Bohr magneton and $g$ is the gyromagnetic ratio) and both uni-axial and transverse anisotropy of magnitude $D$ and  $E$
respectively~\cite{Hir2,Yosida}. The electron-spin interaction Hamiltonian is constructed within the $s$-$d$ model~\cite{Maria,Anna} where the 
transport electron, $s$, are locally exchanged coupled to quantum spins, $\{\mathbf{S}_i\}$ ($d$ indicates that the local moments originating 
from the atomic $d$ shell). In equation (\ref{eq:3}) the interaction strength is $J_\mathrm{sd}$ and $\boldsymbol{\sigma}$ is a 
vector of Pauli matrices. The second term in Eq.~(\ref{eq:3}) represents the potential scattering elastic contribution to the $s$-$d$ interaction 
given by the exchange parameter $J_\mathrm{0}$ (note that this enters as a shift of the on-site potential of a given atom). The ratio, 
$\chi=J_\mathrm{0}/J_\mathrm{sd}$, is typically in the range 1-2 \cite{Delgado1,Delgado2}. This term was 
not included in our previous works \cite{Hurley,HurleyKondo} as it only becomes important for spin-polarized electrodes.

In order to construct an electron-spin interacting self-energy we must first consider the Keldysh~\cite{Keldysh} contour-ordered single-body 
Green's functions (propagators) for both the electronic ($G$) and the spin ($D$) sub-systems in the electron-spin many-body ground 
state $|\rangle$
\begin{equation}
\label{eq:4} [G(\tau,\tau')]_{\sigma\sigma'}=-i{\langle}|T_C\{c_{\sigma}(\tau)c^{\dagger}_{\sigma'}(\tau')\}|{\rangle}\:,
\end{equation}
\begin{align}
\label{eq:5}
[D(\tau,\tau')]_{nm}=-i{\langle}|T_C\{d_{n}(\tau)d^{\dagger}_{m}(\tau')\}|{\rangle}\:.
\end{align}
These propagators describe a non-equilibrium system at zero-temperature. Here $d_m^{\dagger}$ is a quasi-particle creation operator defined 
by the relation
\begin{align}
\label{eq:6}
S^i(\tau)=\sum_{mn}S^i_{mn}d^{\dagger}_m(\tau)d_n(\tau)\:,
\end{align}
where $i=\{x,y,z\}$ and $m,n$ are eigenstates of $H_\mathrm{sp}$ of energy $\varepsilon_{mn}$. The matrix elements 
$S^{i}_{mn}={\langle}m|S^{i}|n{\rangle}$ determine the transition rates from the initial state $n$ to the final state $m$. The quasi-particle 
operators are assumed fermionic in nature\cite{Hurley}. Therefore the equilibrium population of a given state $m$ is given by 
\begin{align}
\label{eq:7}
P_m=d_m^{\dagger}d_m=\frac{1}{\text{e}^{\left(\frac{E-\varepsilon_m}{k_\mathrm{B}T}\right)}+1}\:,
\end{align}
where $E$ is the energy, $T$ the temperature and $k_\mathrm{B}$ the Boltzmann constant.  

In the non-interacting case ($J_\mathrm{sd}=0$) the electronic system is not in equilibrium as the interaction with the electrodes
establishes a steady state current. In contrast the spin system is in thermal equilibrium at the temperature $T$. In this case the energy
resolved lesser and greater Green's functions take the form
\begin{align}
\label{eq:71}
&[G_0^{\lessgtr}(E)]_{\sigma\sigma'}=\frac{[\Sigma^{\lessgtr}_{\mathrm{tip-S}}(E)]_{\sigma\sigma'}+[\Sigma^{\lessgtr}_{\mathrm{sub-S}}(E)]_{\sigma\sigma'}}{(E-\varepsilon_0)^2+\Gamma^2}\:,\\
\label{eq:72}
&[D_0^{\lessgtr}(E)]_{mn}=\frac{[\Pi^{\lessgtr}_{0}(E)]_{mn}+[\Pi_0^{\lessgtr}(E)]_{mn}}{(E-\varepsilon_m)^2+(k_\mathrm{B}T)^2}\:.
\end{align}
In the electronic non-interacting Green's function [see Eq.~(\ref{eq:71})] the coupling of the sample to the tip and the substrate causes a 
broadening of the bare on-site level $\varepsilon_0$ of magnitude $\Gamma=\sum_{\eta\:\sigma\sigma'}[\Gamma_{\eta-S}]_{\sigma\sigma'}$, 
where $\eta=\{\mathrm{tip,sub}\}$. The non-interacting self-energies take the form 
$[\Sigma_{{\eta}}^{>}(E)]_{\sigma\sigma'}=[1-f_{\eta}(E,V)][\Gamma_{{\eta}-S}]_{\sigma\sigma'}$ and 
$[\Sigma_{{\eta}}^{<}(E)]_{\sigma\sigma'}=f_{\eta}(E,V)[\Gamma_{{\eta}-S}]_{\sigma\sigma'}$, 
where $f_\eta(E,V)$ is the Fermi function in each of the $\eta$-th leads at a bias $V$.

In contrast the local spin Green's function of equation (\ref{eq:72}) describes a system, which is adiabatically coupled to a heat-bath of 
temperature $T$. This provides a very weak broadening of the single spin states $\varepsilon_m$ of magnitude $k_\mathrm{B}T$.
Such a heat bath keeps the spin-system in equilibrium and in the non-interacting case the population then resides mostly in the ground 
state [see Eq.~(\ref{eq:7})]. For a ground state population of the spin system $P_m^0$ we have 
$[\Pi_{0}^{>}(E)]_{mn}=\delta_{mn}(1-P^0_m)k_\mathrm{B}T$ and 
$[\Pi_{0}^{<}(E)]_{mn}=\delta_{mn}P^0_mk_\mathrm{B}T$.

\subsection{Spin-polarized electron self-energy}

In order to evaluate the effects that the interaction has on the electronic motion we must calculate the electron-spin self-energy. Here we take a perturbative 
approach and formally expand equation~(\ref{eq:4}) up to the $n$-th order in the interaction Hamiltonian, $H_\mathrm{e-sp}$, as
\begin{align}
\label{eq:8}
&[G(\tau,\tau')]_{\sigma\sigma'}=\sum_n\frac{(-i)^{n+1}}{n!}\int\limits_C{d}\tau_1\dots\int\limits_C{d}\tau_n\ \times \nonumber \\ &\frac{{\langle}0|T_C\{{H}_\mathrm{e-sp}(\tau_1)\dots{H}_\mathrm{e-sp}(\tau_n)c_{\sigma}(\tau)c_{\sigma'}^{\dagger}(\tau')\}|0{\rangle}}{U(-\infty,-\infty)},
\end{align}
where $U$ is the time-evolution unitary operator and the time-averages are performed over the known non-interacting $(J_\mathrm{sd}=0)$ 
ground state $|0{\rangle}$. The time integration over $\tau$ is ordered on the contour $C$ going from $-\infty$  to $+\infty$ and then returning 
from $+\infty$ to $-\infty$, since the ground state of the non-equilibrium system can only be defined at $-\infty$~\cite{Haug}. 

In the following we consider the tip to have a spin polarization $\eta$ ($-1<\eta<1$). This is defined as the spin asymmetry in the electronic coupling 
between the tip and the sample. An such the spin resolved electronic broadening is given by 
$[\Gamma_\mathrm{tip-S}]_{\uparrow\uparrow}=\frac{(1+\eta)}{2}\Gamma_\mathrm{tip-S}$ 
and $[\Gamma_\mathrm{tip-S}]_{\downarrow\downarrow}=\frac{(1-\eta)}{2}\Gamma_\mathrm{tip-S}$, where $\Gamma_\mathrm{tip-S}$ is the 
non-spin-polarized broadening. The substrate is assumed to remain non-magnetic. As a result 
$[G^{\lessgtr}_0(E)]_{\uparrow\uparrow}\neq[G^{\lessgtr}_0(E)]_{\downarrow\downarrow}$ so that we now must retain the spin indexes when constructing
the self-energy. The self-energy for the majority ($\uparrow$) and minority ($\downarrow$) spins writes respectively as 
\begin{align}
\label{eq:10a}
&[\Sigma^{\lessgtr}_\mathrm{int}(E)]_{\uparrow\uparrow}^{(2)}=-J^2_\mathrm{sd}\sum_{mn}[G^{\lessgtr}_0(E\pm\Omega_{mn})]_{\uparrow\uparrow} \times\nonumber \\
&\Big(\delta_{nm}\chi P_nS_{mn}^z+ P_n(1-P_m)|S_{mn}^z|^2\Big)\nonumber \\
&-J^2_\mathrm{sd}\sum_{mn}[G^{\lessgtr}_0(E\pm\Omega_{mn})]_{\downarrow\downarrow}P_n(1-P_m)|S^+_{mn}|^2
\end{align}
and
\begin{align}
\label{eq:10b}
&[\Sigma^{\lessgtr}_\mathrm{int}(E)]_{\downarrow\downarrow}^{(2)}=-J^2_\mathrm{sd}\sum_{mn}[G^{\lessgtr}_0(E\pm\Omega_{mn})]_{\downarrow\downarrow} \times\nonumber \\
&\Big(-\delta_{nm}\chi P_nS_{mn}^z+P_n(1-P_m)|S_{mn}^z|^2\Big)\nonumber \\
&-J^2_\mathrm{sd}\sum_{mn}[G^{\lessgtr}_0(E\pm\Omega_{mn})]_{\uparrow\uparrow}P_n(1-P_m)|S^-_{mn}|^2\:.
\end{align}
The lesser (greater) self-energy describe an incoming (outgoing) electron that excites (relaxes) the spin system by $\Omega_{mn}$, with a probability 
that depends on the occupation of the spin levels $P_m$ and $P_n$ and on the spin selection rules $S^{z,+,-}_{mn}$ 
(note $S^+=S^x+iS^y$ and $S^-=S^x-iS^y$). 
The first term in both the equations (\ref{eq:10a}) and (\ref{eq:10b}), proportional to $\delta_{nm}$, corresponds to the magnetoresistive elastic term of 
the $s$-$d$ Hamiltonian of equation (\ref{eq:3}). The remaining contributions are inelastic in nature and depend on the spin orientation of the electron 
transferred from the tip.

\subsection{Spin-polarized spin self-energy}
\label{SPSE}

When is magnetic and the tunneling current tip carries a finite spin-polarization the spin system can be dragged out of equilibrium, in particular if the current 
density is intense. This means that the equilibrium conditions employed previously~\cite{Hurley, HurleyKondo}, namely $P_{m=\mathrm{GS}}=1$ and 
$P_{m\neq\mathrm{GS}}=0$ is no longer valid. As a consequence we must now derive also an expression for the propagator and thus for the self-energy 
associated to the local spins. The derivation, up to second order in the electron-spin interaction is described in details in the appendix. In particular the 
total spin-self-energy also includes a zeroth-order contribution, which accounts for the non-interacting ($J_\mathrm{sd}=0$) case. This is approximated 
by $[\Pi_0^<(\omega)]_{kk}=P_k^0k_\mathrm{B}T$ and $[\Pi_0^>(\omega)]_{kk}=(1-P_k^0)k_\mathrm{B}T$ where $P_k^0$ is the ground state population. Therefore, in absence of inelastic scattering, the spin system will remain in thermal equilibrium with the heat bath and only the ground state will be 
occupied.

By combining the zeroth and second order contributions to self-energy we can write down a master equation, describing the non-equilibrium spin-population, 
in terms of the total self energy $\Pi^{\lessgtr}(E)$\cite{Mahan}
\begin{align}
\label{eq:16}
\frac{dP_n}{dt}=\frac{1}{\hbar}\sum_{m}\int\limits_{-\infty}^{+\infty}{d}{E}\Big\{&[\Pi^{>}(E)]_{nm}[D_0^<(E)]_{mn}
\nonumber \\
&-\Pi^{<}(E)]_{nm}[D_0^>(E)]_{mn}\Big\}\:.
\end{align}
After some rearrangement this can be written in more compact form as
\begin{align}
\label{eq:17}
\frac{dP_n}{dt}=&\sum_{l}\Big[P_n(1-P_l)W_{ln}-P_l(1-P_n)W_{nl}\Big]\nonumber \\
&+(P_n^0-P_n)k_\mathrm{B}T\:,
\end{align}
where the bias dependent transition rate from an initial state $|l\rangle$ to a final state $|n\rangle$ is calculated after evaluating the integral described in 
 equation (\ref{eq:14}) of the appendix. 
This finally writes
\begin{align}
\label{eq:20}
&W_{nl}=-4\frac{(\rho J_\mathrm{sd})^2}{\Gamma}\sum_{\eta\:\eta'}\zeta(\mu_{\eta}-\mu_{\eta'}+\Omega_{ln})\times
 \nonumber \\
&\Big\{\chi S^z_{nn}\Big([\Gamma_{\eta}]_{\uparrow\uparrow}[\Gamma_{\eta'}]_{\uparrow\uparrow}-[\Gamma_{\eta}]_{\downarrow\downarrow}[\Gamma_{\eta'}]_{\downarrow\downarrow}\Big) +\nonumber \\
&  +|S^z_{nl}|^2[\Gamma_{\eta}]_{\uparrow\uparrow}[\Gamma_{\eta'}]_{\uparrow\uparrow}+|S^z_{nl}|^2[\Gamma_{\eta}]_{\downarrow\downarrow}[\Gamma_{\eta'}]_{\downarrow\downarrow} + \nonumber \\
&  +|S^+_{nl}|^2[\Gamma_{\eta}]_{\downarrow\downarrow}[\Gamma_{\eta'}]_{\uparrow\uparrow}+|S^-_{nl}|^2[\Gamma_{\eta}]_{\uparrow\uparrow}[\Gamma_{\eta'}]_{\downarrow\downarrow}\Big\}\:,
\end{align}
where $\zeta(x)=x/\left(1-e^{-x/k_\mathrm{B}T}\right)$ and $\mu_{\eta}$ is the chemical potential in lead $\eta=\{\mathrm{tip,sub}\}$. Note that 
$\zeta(x)$ is such that for $\eta=\eta'$ the resulting transition rates $W_{nl}$ are bias independent and do not contribute to the current. However, 
they do contribute to the spin relaxation time  i.e. to the time taken by the localized spin system to relax back to its equilibrium state. Such a relaxation 
time is reduced if the coupling between the sample and the leads is increased. Furthermore, the smaller is the inelastic energy transition 
$\Omega_{mn}$, the longer the spins will remain in the excited state before relaxing back to equilibrium. Finally, we note that the above expression
is based on the assumption that the on-site energy is large enough for the density of states of the spin system to remain constant in the small energy 
window of interest. Therefore $\rho=\Gamma/(\varepsilon_0^2+\Gamma^2)$. 

Returning to equation (\ref{eq:17}) note that we are only interested in the steady state non-equilibrium population of the spin states at a given bias. 
Therefore we can set $dP_n(t)/dt=0$ and reduce Eq.~(\ref{eq:17}) to system of linear equations, which can be solved self-consistently.
For an initial guess of the populations ($P_l=P_l^0$) we can iterate Eq.~(\ref{eq:17}), which define $P_n$, with equation (\ref{eq:14}) of the 
appendix, which define the self-energy $\Pi^\lessgtr$, until self-consistency is reached. We can then combine the resulting non-equilibrium population with 
the second order electronic self-energy calculated in Eq.~(\ref{eq:4}) to obtain the current.

\subsection{Real part of the electron self-energy}
In order to provide an explanation to the inherent asymmetry that has been observed in most of the STM experiments on magnetic atoms 
using a non-magnetic tip we return to the expression for the full retarded self-energy. This is defined by the Hilbert transform (note we will 
only consider the non-spin polarized case for simplicity)
\begin{align}
\label{eq:21}
\Sigma_{\mathrm{int}}(E)=\mathcal{PV}\int\limits^{+\infty}_{-\infty}&\frac{dE'}{2\pi}\frac{\Sigma_{\mathrm{int}}^>(E')+\Sigma_{\mathrm{int}}^<(E')}{E-E'} +\\ \nonumber& 
-\frac{i}{2}\{\Sigma_{\mathrm{int}}^>(E')+\Sigma_{\mathrm{int}}^<(E')\}\:.
\end{align}
By using the expressions derived in Section~\ref{SPSE} for the 2$^\mathrm{nd}$ order lesser and greater self-energies we can find an analytic 
expression for the real contribution to the retarded self energy
\begin{align}
&\label{eq:22}
\mathrm{Re}[\Sigma_{\mathrm{int}}(E)^{(2)}]=2\rho J^2_\mathrm{sd}\sum_{i,m,n}|S^i_{mn}|^2P_n(1-P_m)\times \\ \nonumber& 
\frac{1}{\Gamma}\Big\{2\pi\varepsilon_0+\sum_{\eta}\Gamma_{\mathrm{\eta-S}}\:\mathrm{ln}\Big[\frac{(E+\Omega_{mn}-\mu_{\eta})^2+(k_BT)^2}{(E-\Omega_{mn}-\mu_{\eta})^2+(k_BT)^2}\Big]\Big\}\:.
\end{align}
Such a final expression is heavily dependent on the on-site energy $\varepsilon_0$ but is also an odd function of the energy and the bias via its 
logarithmic dependence on the spin level eigenvalues with opposite polarity for $+\Omega_{mn}$ and $-\Omega_{mn}$. We will show in the results 
section that this is at the origin of the conductance asymmetry found in experiments.

\subsection{Electronic current}
We can finally unveil the effects that a non-equilibrium spin-population bares on the conductance profile of a magnetic nanostructure by taking the 
derivative of the current with respect to the bias voltage $V$. The current, $I_{\eta}$, flowing at the electrode ${\eta}=\{\mathrm{tip, sub}\}$ can be
written as
\begin{align}
\label{eq:23}
&I_{\eta}=\int_{-\infty}^{+\infty}dE\bar{I_{\eta}}(E)\:,\\
\label{eq:24}
\bar{I_{\eta}}(E)=\frac{q}{h}\text{Tr}\{&[\Sigma_{\eta}^{>}(E)G^{<}(E)]-[\Sigma_{\eta}^{<}(E)G^{>}(E)]\}\:,
\end{align}
where $G^{\lessgtr}(E)$ are the full many-body lesser/greater electronic Green's functions. These are finally defined as
\begin{align}
\label{eq:73}
&[G^{\lessgtr}(E)]=\frac{[\Sigma^{\lessgtr}_{\mathrm{tip-S}}(E)]+[\Sigma^{\lessgtr}_{\mathrm{sub-S}}(E)]+[\Sigma^{\lessgtr}_{\mathrm{int}}(E)]}{(E-\varepsilon_0-\mathrm{Re}[\Sigma_{\mathrm{int}}(E)])^2+(\Gamma-\mathrm{Im}[\Sigma_{\mathrm{int}}(E)])^2},
\end{align}
and $[\Sigma^{\lessgtr}_{\mathrm{int}}(E)]=[\Sigma^{\lessgtr}_{\mathrm{int}}(E)]^{(2)}+[\Sigma^{\lessgtr}_{\mathrm{int}}(E)]^{(3)}$ where the third order self-energies 
are calculated following Ref.~[\onlinecite{HurleyKondo}]. Here for simplicity we take the expression for the third order contribution to $\Sigma_\mathrm{int}$ obtained by neglecting any explicit spin-polarization. 
Such approximation is justified by the fact that the effects due to spin-polarization are small at the third order and that in doing so we avoid a rather cumbersome formulation. We do however consider the combination of non-spin polarized 3rd order effects with 2nd order spin-polarized self energies to highlight subtle differences in the spectra.

\section{Results}
\subsection{Non-equilibrium population}
We start our analysis by first looking at the effects originating from driving the spin system out of equilibrium with an electronic current. This attempts at explaining 
the experiments reported in Ref.~[\onlinecite{Loth1}], in which a STM tip (non-magnetic) is positioned above a Mn dimer deposited onto a CuN substrate. The conductance 
spectra are measured for different tip to sample distances. Varying the STM tip height is equivalent to changing both the current density and the electronic coupling between 
the tip and the sample. Non-equilibrium effects then appear as variations of the conductance profiles as a function of the STM tip height.

\begin{figure}[h]
\centering
\resizebox{\columnwidth}{!}{\includegraphics[width=2cm,angle=-90]{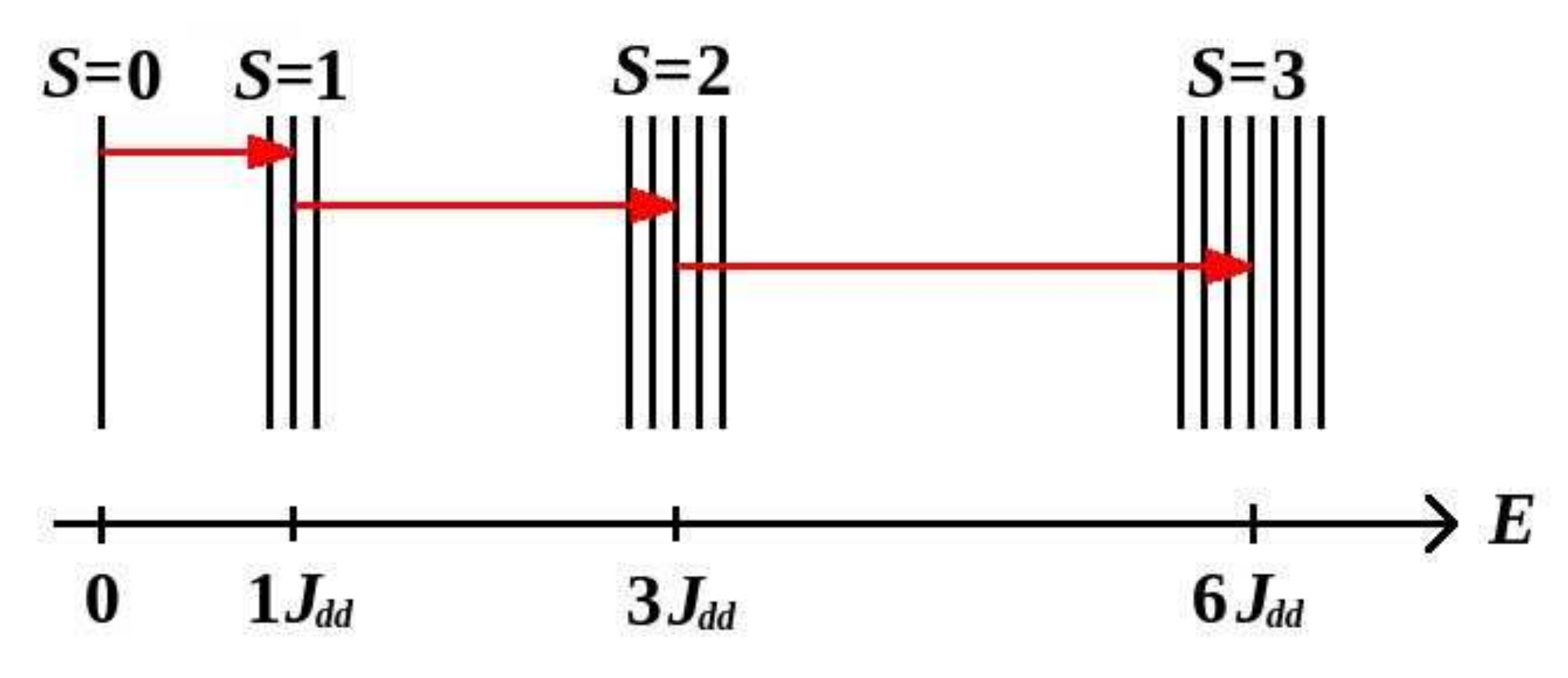}}
\caption{Excitation spectrum for an antiferromagnetically exchanged coupled Mn dimer deposited on CuN. The ground state is a $S=0$ spin singlet. 
The first three excited state multiplets have respectively spin $S=1$, $S=2$ and $S=3$. In the figure we also indicate the energy separation between the various 
spin multiplets in units of the exchange parameter $J_\mathrm{dd}$.}
\label{Fig1}
\end{figure}
Many of the parameters needed by our model can be extrapolated from a similar experiment carried on over Mn linear atomic chains deposited on CuN~\cite{Hir1}. 
The five unpaired electrons in the Mn 3$d$ shell suggest a $S=5/2$ ground state, as confirmed both by experiments and theory~\cite{Lorente,Novaes}. 
The spin-spin exchange interaction between two Mn atoms is antiferromagnetic and has an estimated value of $J_\mathrm{dd}=6.2$~meV. As a result the ground 
state of the dimer is a singlet (total spin $S=0$). The first excited state is a triplet with total spin $S=1$ and the energy splitting between the ground state and such 
first excited state is exactly $J_\mathrm{dd}$. The next excited level is the quintuplet with total spin $S=2$ and it is separated from the first excited state by 
$2J_\mathrm{dd}$. This pattern continues throughout the spin manifold (see figure~\ref{Fig1}). The axial and transverse anisotropies are found to be $D=-0.037$~meV 
and $E=0.007$~meV respectively. These cause the lifting of the spin multiplets degeneracy. The temperature is set at $T=0.5$~K. The value of $J_\mathrm{sd}$ is 
estimated from density functional theory (DFT) to be of the order of 500~meV~\cite{Lucignano}, while $\Gamma_\mathrm{sub-S}$ is also found from DFT to be 
approximately 100~meV. In contrast $\Gamma_\mathrm{tip-S}$ remains an adjustable parameter with the chosen values ranging from 0.125~meV to 200~meV. Finally, 
in order to ensure a nearly constant density of states around the Fermi energy ($E_F=0$) we set the on-site energy of the atom under the STM tip to be $\varepsilon_0=1$~eV.

\begin{figure}[h]
\centering
\resizebox{\columnwidth}{!}{\includegraphics[width=2cm,angle=0]{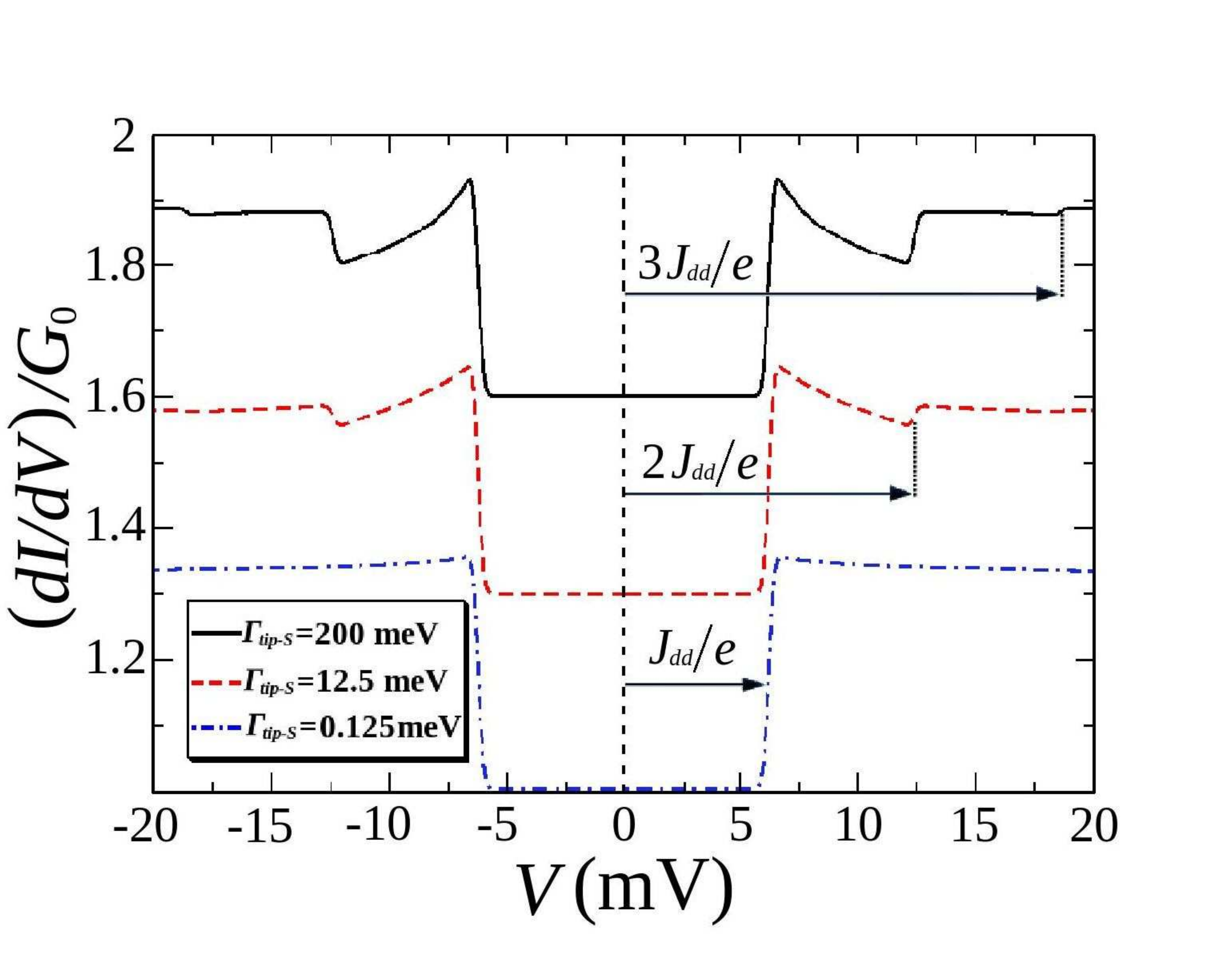}}
\caption{(Color online) Normalized conductance spectra for the Mn dimer calculated at different tip to sample distance, i.e. for different $\Gamma_{\mathrm{tip-S}}$ coupling
strengths. We notice that the stronger is the coupling, the more the system is driven out of equilibrium. This results in the appearance of additional spin transitions,
which manifest themselves as steps or drops of the conductance as a function of bias.}
\label{Fig2}
\end{figure}
Figure~\ref{Fig2} shows the conductance spectra obtained by simply taking the numerical derivative of the current [Eq.~(\ref{eq:23})] with respect to the bias. We consider
three different tip to sample distance, corresponding respectively to weak ($\Gamma_\mathrm{tip-S}=0.5$~meV), intermediate ($\Gamma_\mathrm{tip-S}=50$~meV) and 
strong ($\Gamma_\mathrm{tip-S}=200$~meV) electronic coupling. The evolution of the conductance lineshape as a function of $\Gamma_\mathrm{tip-S}$ is a direct 
consequence of the spin system being driven out of equilibrium. For $\Gamma_\mathrm{tip-S}=0.5$~meV the STM tip is far enough from the sample to ensure 
that the spin system is always in its ground state between two subsequent electron tunneling events. Therefore the only transition detected in the $G(V)$ profile is
that between the $S=0$ ground state and the first excited state with $S=1$. This has an excitation energy equal to $J_\mathrm{dd}$ and it does manifest itself as a 
conductance step at a voltage $V=J_\mathrm{dd}/e$, with $e$ being the electron charge. 

As the tip is brought closer to the sample ($\Gamma_\mathrm{tip-S}=50$~meV) the first excited triplet level ($S=1$) starts to populate. Now an incoming electron 
with sufficiently large energy ($2J_\mathrm{dd}$) can induce a second transition from from the first to the second excited state. Note that the $S=2$ state is not accessible 
with a single electron tunneling process from the ground state and it can be reached only if the spin system does not have enough time between tunneling events to relax
back to the ground state. For this case the transition appears as a reduction of the conductance at the critical voltage $V=2J_\mathrm{dd}/e$. The same spectroscopical 
feature is further enhanced at an even larger current density ($\Gamma_\mathrm{tip-S}=$~200~meV), when a third conductance step appears at $3J_\mathrm{dd}/e$. This
is associated to a transition from the $S=2$ to the $S=3$ spin state and it becomes possible only if the occupation of the $S=2$ level is not zero, i.e. if the system
is driven to this highly excited state. These results are in almost perfect quantitative agreement with the experimental data (see Fig.~2 of Ref.~[\onlinecite{Loth1}]).

\begin{figure}[h]
\centering
\resizebox{\columnwidth}{!}{\includegraphics[width=5cm,angle=0]{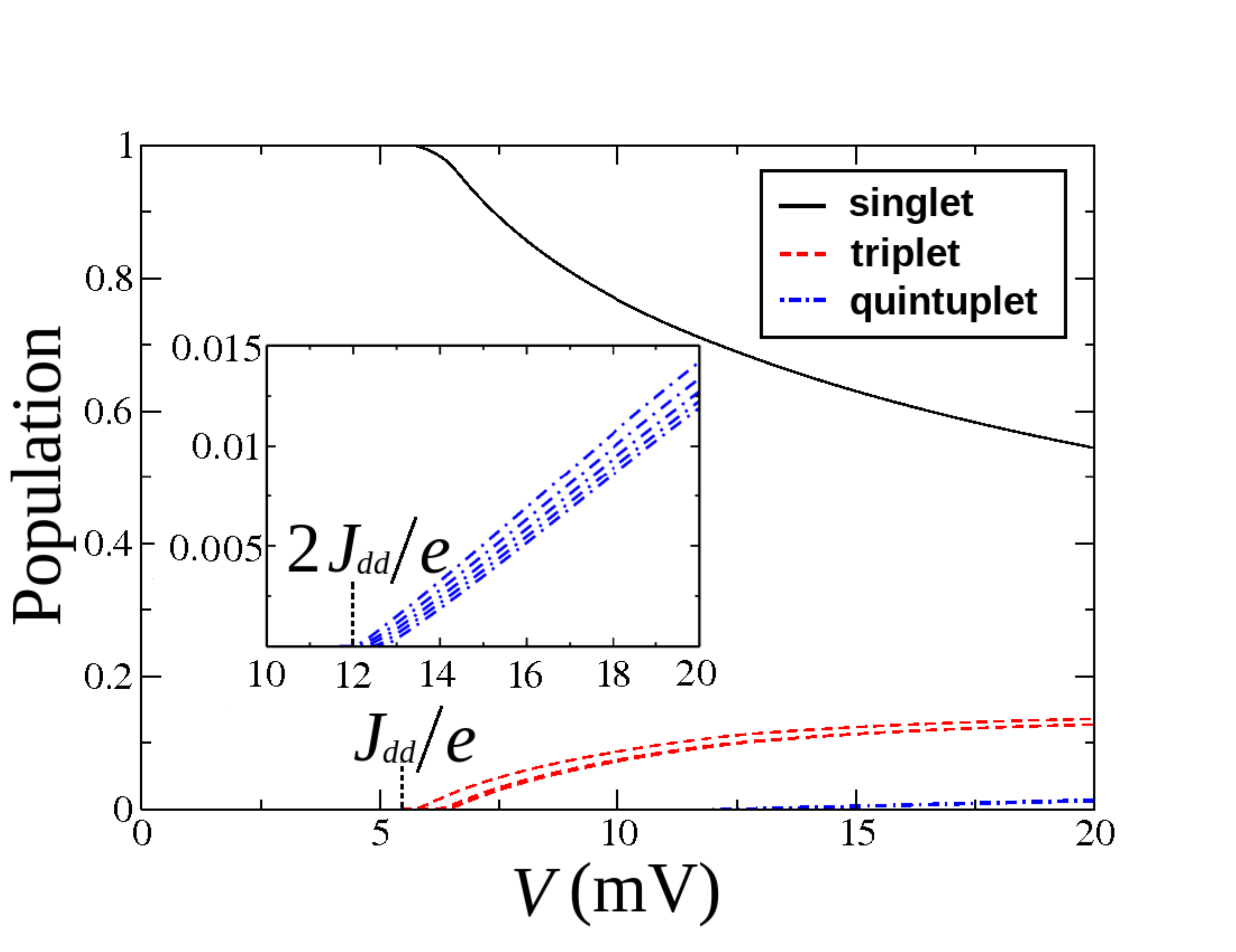}}
\caption{(Color online) Non-equilibrium population of the Mn dimer singlet ($S=0$), triplet ($S=1$) and quintuplet ($S=2$) states. The inset shows a magnified view of 
the population of the $S=2$ state as it start to get populated at approximately 12~mV.}
\label{Fig3}
\end{figure}
The evolution of the population of the various spin states (up to $S=2$) as a function of bias is presented in figure~\ref{Fig3}. This is calculated in the case of strong
tip to sample electronic coupling $\Gamma_\mathrm{tip-S}=$~200~meV. In the figure one can note the strong spin-pumping from the ground state into both the first 
and the second excited state. The excitation to the $3^\mathrm{rd}$ excited state occurs at  approximately 18~meV but is too weak to be observed on this scale.

\subsection{Spin-Polarized tip}

We now move on to consider the situation where the tip is magnetic, i.e. when the injected current is spin-polarized. Again we use as guide the experimental work 
of Loth {\it et al.}\cite{Loth1,Loth2}. The STM tip is now spin-polarized by placing an additional Mn atom at its apex, while also applying a strong magnetic field  perpendicular to 
the substrate of 3~T. In this case the spectrum is collected from a single Mn or Fe ion on the surface (not from a dimer). In the case of Mn, the atom exhibits a weak anisotropy on 
CuN (see previous section) and the strong magnetic field effectively produces a Zeeman split of the six levels of the $S=5/2$ Mn spin manifold. The direction of the magnetic 
field in these experiments is chosen so that the ground state of the Mn spin corresponds to the magnetic quantum number $m=+5/2$. Since the same magnetic field is 
applied to the Mn atom on the tip's apex, the tip and atom are both spin-polarized and collinear.
\begin{figure}[h]
\centering
\resizebox{\columnwidth}{!}{\includegraphics[width=5cm,angle=-90]{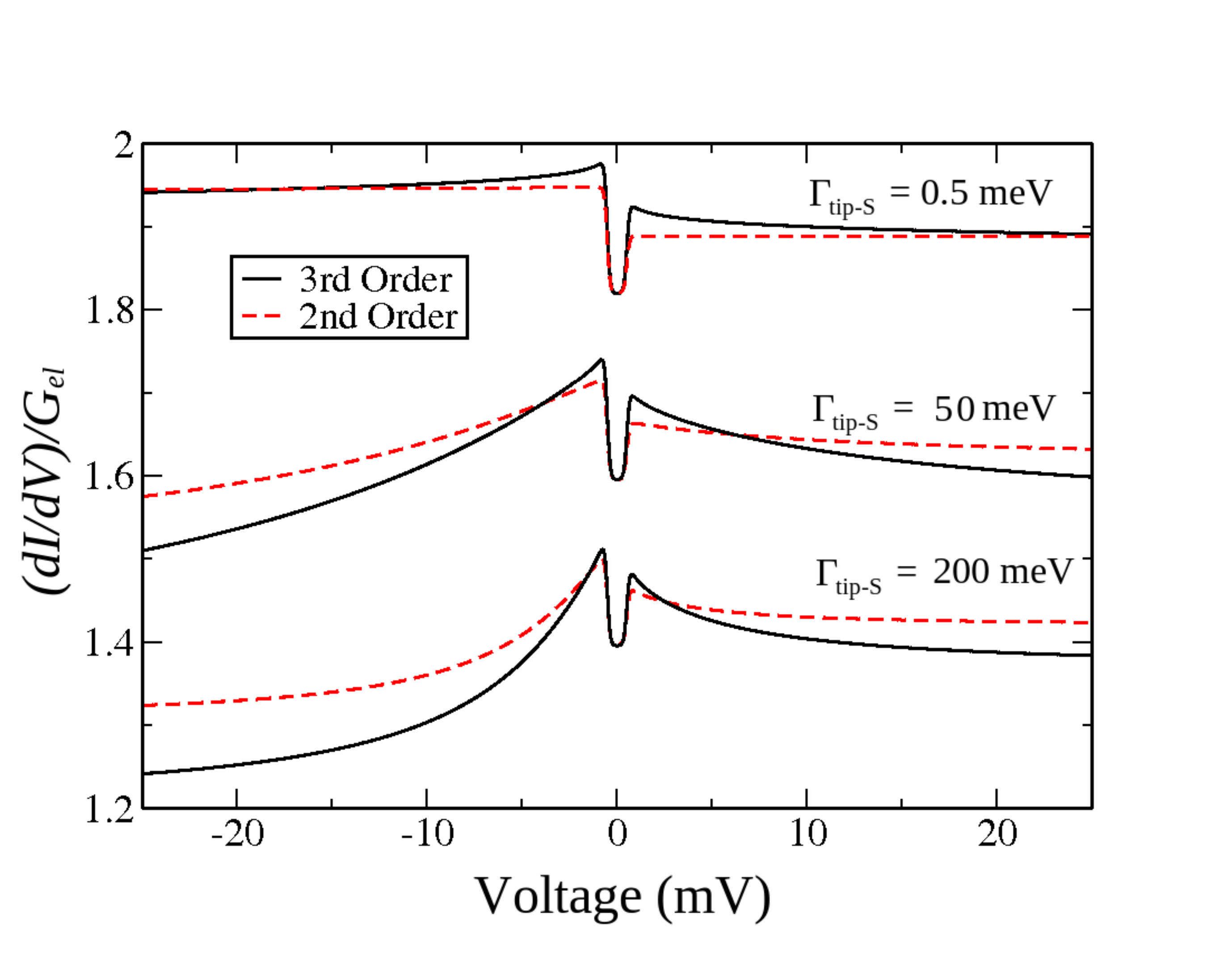}}
\caption{(Color online) Normalized conductance spectra for a single Mn atom explored by STM at different tip to sample electronic couplings $\Gamma_\mathrm{tip-S}$.
Here a magnetic field of 3~T is applied along the z-axis. The asymmetry in the conductance profile is is due to the spin-polarization of the tip. Such an 
asymmetry is more pronounced as the system is driven further away from equilibrium. Comparison is made between second and third order calculations.}
\label{Fig4}
\end{figure}

Figure~\ref{Fig4} shows the calculated spectra for the system described above. In particular we consider magnetic field strength of 3~T and either 
weak ($\Gamma_{\mathrm{tip-S}}=0.5$~meV), intermediate ($\Gamma_{\mathrm{tip-S}}=50$~meV) or strong ($\Gamma_{\mathrm{tip-S}}=200$~meV) 
tip to sample couplings. The on-site energy is fixed at $\varepsilon_0=2$eV and the value of $J_{\mathrm{sd}}=500$meV is infered from the work of 
Lucignano~\emph{et al.}\cite{Lucignano}. The tip spin-polarization constant and the inelastic ratio that best fit the experimental data 
are respectively $\eta=-0.3$ and $\chi=1.5$. In the weak coupling regime (when the local spin remains always close to equilibrium) the local spin resides 
almost entirely in its $m=+5/2$ ground state. Due to the spin-exchange selection rules and to the collinearity of the tip and the sample, only the minority carriers 
can excite the local spin out of the ground state. For a tip spin-polarization of $\eta=-0.3$, there are more minority electrons coming from the tip than those 
coming from the substrate. As a result, the intensity of the inelastic interaction will change depending on the direction of the current. This creates an asymmetry 
in the conductance spectrum with respect to the applied bias. The additional lineshape features appearing in the weak coupling case (the conductance decay
following a conductance step) are due to the third order Kondo-like self-energy, which produces a logarithmic decay at the conductance steps. This result is 
in good agreement with experiments (see Fig.~4 of Ref.~\onlinecite{Loth1}).

\begin{figure}[t]
\centering
{\resizebox{\columnwidth}{!}{\includegraphics[width=5cm,angle=-90]{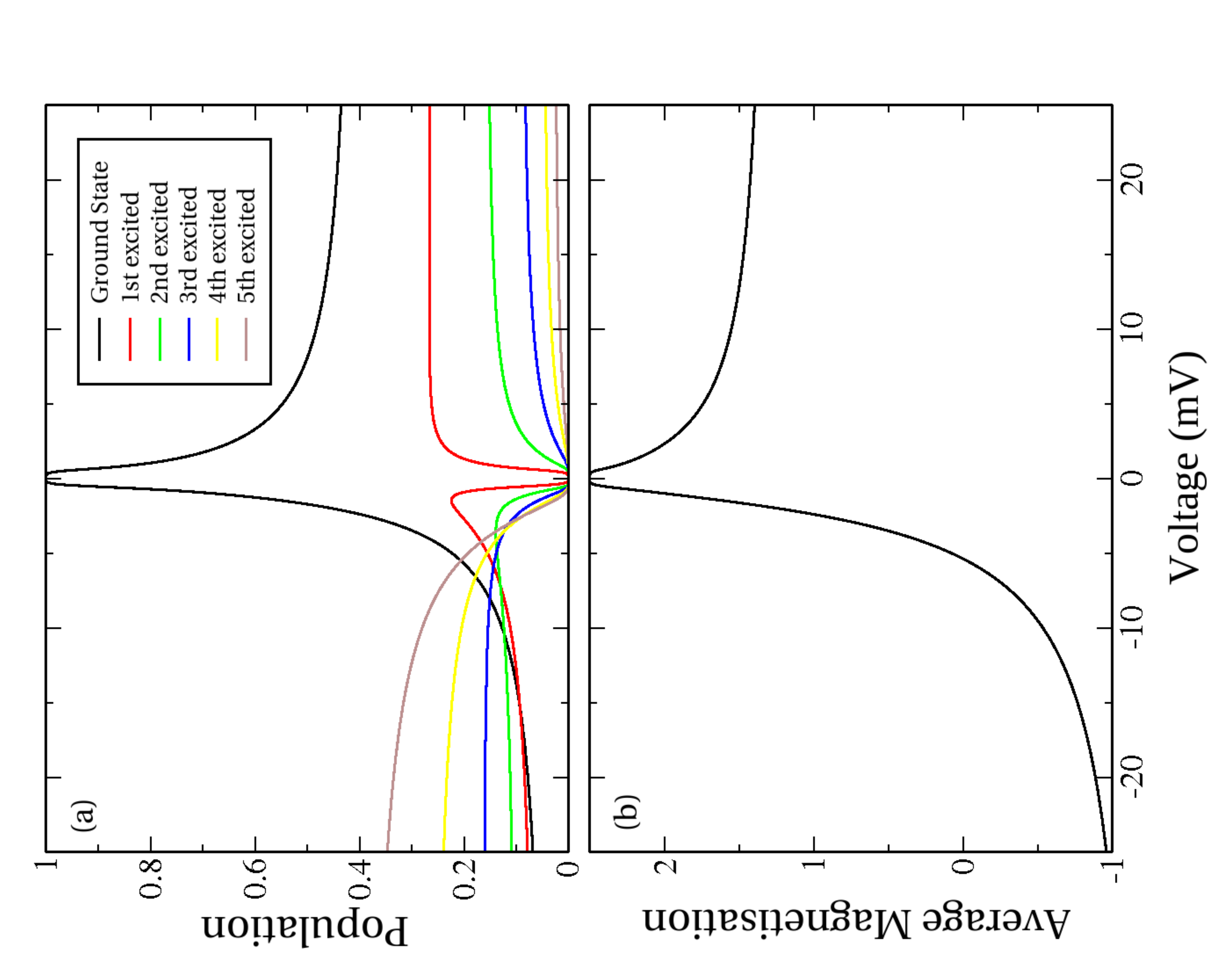}}
\caption{(Color online) (a) Non-equilibrium population of the various spin states of Mn on CuN as a function of bias and (b) the resulting average 
magnetization. These have been calculate for a magnetic field of 7~T aligned along the $z$-axis and for strong tip to sample electronic coupling 
$\Gamma_{\mathrm{tip-S}}=200$~meV. We notice that as the spin is driven far away from its equilibrium ground state the magnetization flips
its direction.}}
\label{Fig5}
\end{figure}

When the spin of the Mn ion is driven further out of equilibrium, in particular in the strong coupling case, the bias asymmetry becomes more pronounced. 
such spin-pumping phenomenon can be appreciated by looking at figure~\ref{Fig5}(A), where we show the populations of the six spin states of the Mn atom 
as a function of bias for strong tip to sample coupling ($\Gamma_{\mathrm{tip-S}}=200$~meV). From the figure one can see that as the bias increases the 
$m=+5/2$ ground state gets depleted in favour of populating the other five excited states. In particular already at $V\sim10$~mV the population of the
$m=-5/2$ level is larger than that of the ground state. 
In th figure we also plot the average magnetization, which is defined as $\langle S^z\rangle=\sum_mP_mS^z_{mm}$ [see panel (b)]. Intriguingly we find
that for negative biases the spin is effectively flipped from $m=+5/2$ to $m=-3/4$ over 25meV range. Such spin flipping results in a large dip in the conductance 
for negative biases as the tip is no longer collinear to the sample.

\begin{figure}[t]
\centering
\resizebox{\columnwidth}{!}{\includegraphics[width=5cm,angle=-90]{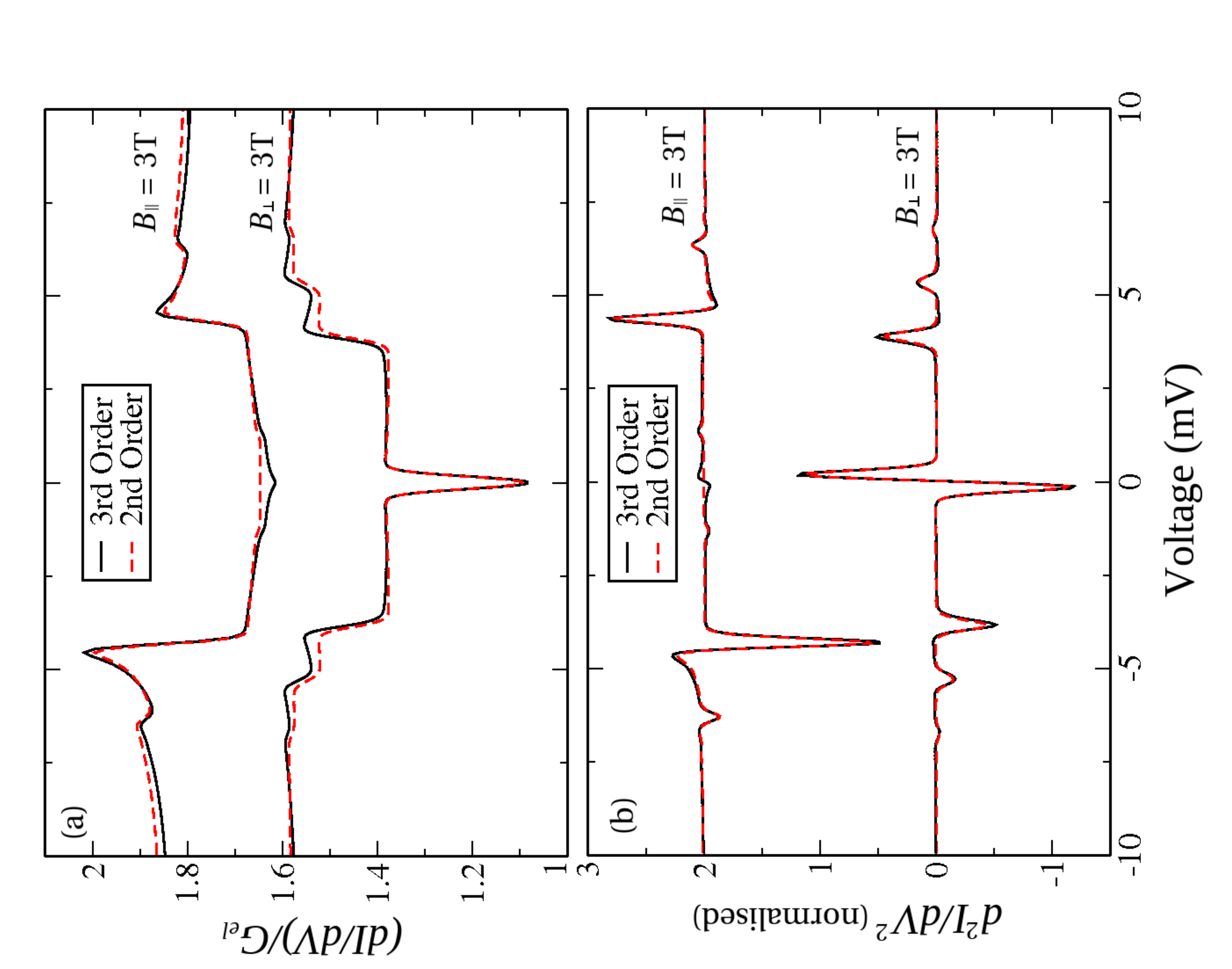}}
\caption{\footnotesize{(Color online) (a) 2nd and 3rd order conductance spectra for the Fe atom with spin polarization $\eta=0.35$. The magnetic field of 3T is applied both parallel and perpendicular to the easy axis of the Fe atom. (b) Second dervative of the current for the same spectra shown in (a).}}
\label{Fig6}
\end{figure}

In Fig \ref{Fig6} we present the calculated spectra for the Fe atom in the spin polarized case. We choose parameters in this case that conform with the experimental data of Loth \emph{et al.} in Ref. \cite{Loth2}. The Fe atom is assumed to carry a quantum mechanical spin of $S=2$ and it also exhibits a transverse easy axis anisotopy of $D=-1.53$meV and an axial anisotropy of $E=0.31$meV. We again assume a large value of the onsite energy, $2$eV, and we examine the spectra in the strong coupling case of $\Gamma_{\mathrm{tip-S}}=200$meV with a tip polarisation of $\eta=0.35$ and magnetic field strength 3T as used in experiments. In Fig 6(a) we present the conductance spectra for the two cases when the magnetic field is parallel and perpendicular to the easy axis of the atom (the $z-$axis in this model). As previously, we present this for both second and third order calculations. Firstly, we notice that the spin polarized tip affects the spectra only in the case of parallel magnetic field where a clear bias assymetry is produced. No significant assymetry is found in the perpendicular case. This conforms with the experimental findings and is due to the fact that electron spins in the tip are no longer colinear with localised spin of the Fe atom.

As found in previous works \cite{HurleyKondo} the inclusion of third order effects is vital in reproducing the corresct logarithmic decay at each of the conductance steps, which is particularly noticable for the perpendicular magnetic field. More significantly, experimental spectra for the parallel case exhibit a zero bias conductance dip which is absent in the 2nd order spectra but appears strongly when third order effects are included. This can also be seen the calculation of the second derivative of the current in Fig 6(b) where a clear zero bias anomaly is evident in the third order case.

\subsection{Non-spin polarized asymmetry}

We finally move to discuss the inherent asymmetry measured in the conductance profile, which is usually observed even if the tip is not spin-polarized~\cite{Loth1,Hir2}. 
We model this lineshape feature by including the real part of the full interacting electron-spin self-energy in the description [see equation (\ref{eq:22})]. 
The structure of this contribution to the self-energy shows an explicit dependence on the on-site energy, $\varepsilon_0$, and also a logarithmic peak of 
width $k_\mathrm{B}T$ at the onset of an inelastic transition ($E-\mu_{\eta}=\pm\Omega_{mn}$). The asymmetry arises from the difference in polarity of 
the logarithmic peak for $\pm\Omega_{mn}$. The self-energy is thus an odd function of both energy and bias. This results in the conductance profile having
a decrease of the step heights for $V=-\Omega_{mn}/e$ and a increase of them for $V=+\Omega_{mn}/e$.

We test this approach by considering the case of a non-spin-polarized tip and a single Mn atom. We use the same anisotropy parameters as for the 
Mn dimer but, for the sale of simplicity, we keep the spin always its equilibrium state and choose $\Gamma_{\mathrm{tip-S}}=0.5$meV. 
Figure~\ref{Fig7} shows the resulting conductance spectra for three different choices of the on-site energy $\varepsilon$. It is cear that the closer $\varepsilon$ 
is to the Fermi energy ({0~eV}), the greater is the bias asymmetry, while as  $\varepsilon$ is increased, the conductance profile becomes 
more symmetric. In this respect, the formalism outlined here is in agreement with the Fano lineshape argument~\cite{Fano} where the degree of asymmetry 
for electrons tunneling through a single impurity is given by a ratio of the real to the imaginary contributions to the interacting Green's function\cite{Ujsaghy}.
\begin{figure}[h]
\centering
\resizebox{\columnwidth}{!}{\includegraphics[width=5cm,angle=-90]{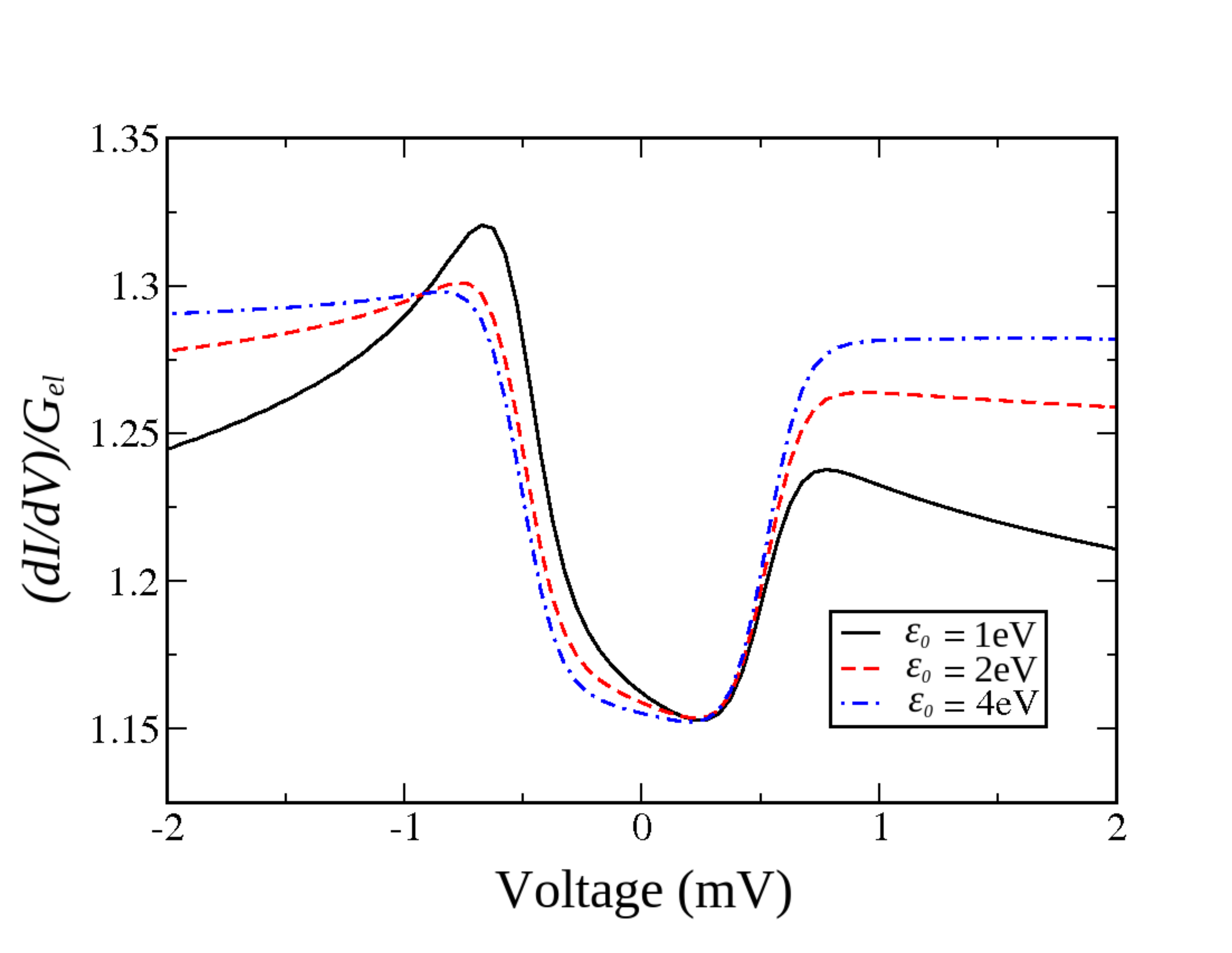}}
\caption{(Color online) Normalised conductance spectra for a single Mn single calculated by including the real contribution to the interacting electron-spin self-energy 
for different on-site energies $\varepsilon_0$. Note that the conductance asymmetry increases with decreasing $\varepsilon_0$, i.e. as the onsite energy moves closer
to the Fermi level.}
\label{Fig7}
\end{figure}

As an additional test we consider the case of a Mn trimer, whose spectrum was shown first by Hijibehedin \emph{et al.}\cite{Hir1} to exhibit a large bias asymmetry 
when measured with a non-magnetic tip. We model this system by choosing an antiferromagnetic nearest neighbour exchange coupling 
$J^{(1)}_{\mathrm{dd}}=$2.3~meV. Furthermore, in order to accurately describe the position of the principle conductance steps in the conductance profile, we also 
include a ferromagnetic second-nearest-neighbour interaction between the local spins of magnitude $J^{(2)}_{\mathrm{dd}}=$-1.0~meV~\cite{Fernandez-Rossier}. 

We again choose to keep the spin system in equilibrium and therefore consider weak coupling between the STM tip and the second atom in the trimer chain 
($\Gamma_{\mathrm{tip-S}}=0.5$~meV). The best fit to the experimental data is found with $\varepsilon_0=-1$eV. Figure~\ref{Fig8} shows the model fit to the 
experimental data (from Ref.~[\onlinecite{Hir1}])). Whereas previous calculations did not predict any conductance asymmetry \cite{Hurley} it is clear from 
the figure that the inclusion of the real part of the self-energy in the description produces a significant conductance asymmetry. This is most prominent at the 
principle step height $(\approx\pm16.5\mathrm{meV})$ for each bias polarity. Although the step height for the negative bias is not as small as that found experimentally, 
the qualitative trends are similar. In particular we notice the logarithmic conductance increase (reduction) that occurs before (after) the onset of the step at 
$V=+16.5$meV, which also originates from the third order contribution to the self-energy.

In this work, based on a perturbative approach of the s-d model, we have shown that the entire lineshape description can be re-conciliated with experiments by considering an expansion of the self-energy to the third order, which also includes its real part. As such we have shown that the conductance asymmetry can be described also if the electronic orbitals forming the sample’s spin are not explicitly taken into account as also suggested by Delgado and Fernandez-Rossier \cite{Delgado2}.
\begin{figure}[h]
\centering
\resizebox{\columnwidth}{!}{\includegraphics[width=5cm,angle=-90]{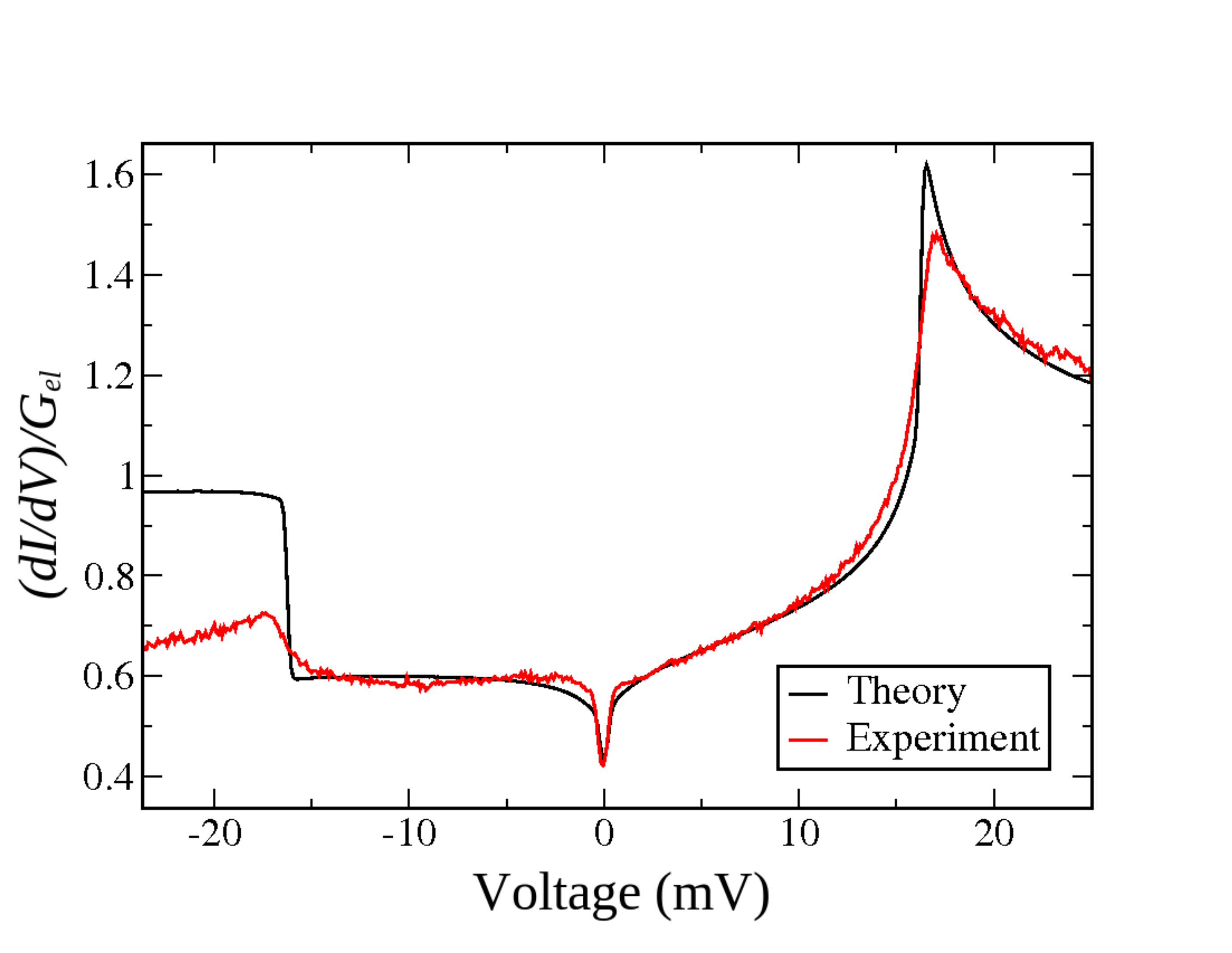}}
\caption{(Color online) Comparison between the experimental (red) and theoretical (black) conductance spectra for a Mn trimer on CuN probed by a non-magnetic
STM tip. In the calculation we have included the real part of the interacting electron-spin self-energy. This provides the conductance asymmetry with bias.}
\label{Fig7}
\end{figure}

\section{Conclusions}
In conclusion we have studied the lineshape details of the conductance profile of Mn atoms deposited on CuN and probed with a STM 
tip either or not carrying spin-polarization. In particular we have looked closely at the asymmetry of the conductance with the bias polarity. 
Firstly, we have extended our perturbative approach to spin-scattering to the spin-polarized case and considered an expansion of the 
complex part of the electronic propagator up to the third order. This allows us to reproduce the logarithmic decay of the conductance 
subsequent a conductance step, which is observed in experiments but could not be explained by a second order theory.

When the current density is increased and the tip is spin-polarized the conductance profile starts to develop a significant asymmetry with respect to
the bias polarity. These are indicative of the spin system being driven out of equilibrium. We have then derived a second order expansion of the
spin-propagator capable of evaluating the non-equilibrium population of the various spin energy levels. This was put favorably to the test against a 
series of experiments probing a single Mn and Fe ions with a spin-polarized STM tip in an intense magnetic field. Furthermore the same formalism was 
capable of describing excitations occurring away from the ground state for a Mn dimer probed by a non-magnetic tip. Also in this case the
agreement with experiments is very satisfactory.

Finally, in an attempt to describe the bias asymmetry in the case of non-spin-polarized STM tips we have derived an analytic expression for the real 
part of the electron-spin interacting self-energy. This contains logarithmic peaks at the excitation energies that are odd with respect to energy and 
voltage. Such parity results in an asymmetry in the conductance profiles. Such a scheme was tested for the case of a Mn monomer and a Mn trimer 
and compares reasonably well with experiments.

\section{Acknowledgments}
This work is sponsored by the Irish Research Council for Science, Engineering \& Technology (IRCSET). NB and SS thank Science 
Foundation of Ireland (grant No. 08/ERA/I1759) and CRANN for financial support. Computational resources have been provided by 
the Trinity Centre for High Performance Computing (TCHPC). We wish to thank Cyrus Hirjibehedin for making the 
experimental data shown in figures \ref{Fig7} available to us.

\section{Appendix}

Here we wish to derive a method for calculating the steady-state non-equilibrium distribution of the spin energy levels populations, $P_m(V)$, due to the coupling 
with the electrodes. In order to do so we expand equation (\ref{eq:5}) up to the $n$-th order in the interaction Hamiltonian
\begin{align}
\label{eq:12}
&[D(\tau,\tau')]_{nm}=\sum_n\frac{(-i)^{n+1}}{n!}\int\limits_C{d}\tau_1\dots\int\limits_C{d}\tau_n\ \times \nonumber \\ &\frac{{\langle}0|
T_C\{{H}_\mathrm{e-sp}(\tau_1)\dots{H}_\mathrm{e-sp}(\tau_n)d_{n}(\tau)d_{m}^{\dagger}(\tau')\}|0{\rangle}}{U(-\infty,-\infty)}\:,
\end{align}
where $U$ is the time-evolution unitary operator and the time-averages are over the known non-interacting $(J_\mathrm{sd}=0)$ ground state 
$|0{\rangle}$. As in equation (\ref{eq:8}) the time integration over $\tau$ is ordered on the contour $C$ going from $-\infty$  to $+\infty$ and then returning 
from $+\infty$ to $-\infty$ \cite{Haug}.

By inserting the expression for $H_\mathrm{e-sp}$ from equation (\ref{eq:3}) into the equation above and by expanding up to the second order we obtain [note 
for the ease of the description we omit the elastic contribution of $J_0$, which is then included in the final expression in equation (\ref{eq:12})]
\begin{align}
\label{eq:9}
&[D(\tau,\tau')]^{(2)}_{nm}=\frac{(-i)^{3}}{2!}J^2_\mathrm{sd}\sum_{\alpha,\alpha',\beta,\beta'}\int\limits_C{d}\tau_1\int\limits_C{d}\tau_2\ \times \nonumber \\ &{\langle}0|T_C\{c_{\alpha}^{\dagger}(\tau_1)c_{\alpha'}(\tau_1)c_{\beta}^{\dagger}(\tau_2)c_{\beta'}(\tau_2)d_{n}(\tau)d_{m}^{\dagger}(\tau')\}|0{\rangle} \nonumber \\
&\times \sum_{i,j}{\langle}0|T_C\{{S^{i}}(\tau_1){S^{j}}(\tau_2)\}|0{\rangle}[{\sigma}^i]_{{\alpha}{\alpha'}}[{\sigma}^j]_{{\beta}{\beta'}}\:,
\end{align}
where the indices $i$ and $j$ run over the cartesian coordinates $x$, $y$ and $z$ for the given spin coupled to the tip (the tip make electronic contact with
one spin only). We now substitute into equation (\ref{eq:9}) the operator breakdown of the spin from equation (\ref{eq:6})
\begin{align}
\label{eq:10}
&[D(\tau,\tau')]^{(2)}_{nm}=\frac{(-i)^{3}}{2!}J^2_\mathrm{sd}\sum_{k,k',l,l'}\int\limits_C{d}\tau_1\int\limits_C{d}\tau_2 \nonumber \\
&\times {\langle}0|T_C\{d_{n}(\tau)d_{k}^{\dagger}(\tau_1)d_{k'}(\tau_1)d_{l}^{\dagger}(\tau_2)d_{l'}(\tau_2)d_{m}^{\dagger}(\tau')\}|0{\rangle} \nonumber \\
&\times \sum_{\alpha,\alpha',\beta,\beta'}{\langle}0|T_C\{c_{\alpha'}(\tau_1)c_{\beta}^{\dagger}(\tau_2)c_{\beta'}(\tau_2)c_{\alpha}^{\dagger}(\tau_1)\}|0{\rangle} \nonumber \\
&\times
\sum_{i,j}S_{kk'}^iS_{ll'}^j[{\sigma}^i]_{{\alpha}{\alpha'}}[{\sigma}^j]_{{\beta}{\beta'}}\:.
\end{align}
The time-ordered contractions of the two brackets in equation (\ref{eq:10}) can be re-written in terms of their respective non-interacting Green's functions, 
$D_0(\tau,\tau')$ and $G_0(\tau,\tau')$ as follows
\begin{align}
\label{eq:11}
&[D(\tau,\tau')]^{(2)}_{nm}=-J^2_\mathrm{sd}\sum_{k,k',l,l'}\int\limits_C{d}\tau_1\int\limits_C{d}\tau_2 \nonumber \\
&\times \delta_{nk}\delta_{lk'}\delta_{ml'}[D_0(\tau,\tau_1)]_{nn}[D_0(\tau_1,\tau_2)]_{ll}[D_0(\tau_2,\tau')]_{mm} \nonumber \\
&\times \sum_{\alpha,\alpha',\beta,\beta'}\delta_{\alpha'\beta}\delta_{\alpha\beta'}[G_0(\tau_1,\tau_2)]_{\beta\beta}[G_0(\tau_2,\tau_1)]_{\alpha\alpha} \nonumber \\
&\times
\sum_{i,j}S_{kk'}^iS_{ll'}^j[{\sigma}^i]_{{\alpha}{\alpha'}}[{\sigma}^j]_{{\beta}{\beta'}},
\end{align}
where the extra factor of 2 emerges from the fact that a second contraction of the time-ordered bracket merely exchanges $\tau_1$ and $\tau_2$. Then, by using 
Dyson's equation~\cite{Mahan}, one can write the second order contribution to the interacting spin self-energy ($\Pi$). This reads
\begin{align}
\label{eq:12}
&[\Pi(\tau_1,\tau_2)]^{(2)}_{nm}=-2J^2_\mathrm{sd}\sum_{\alpha,\beta}[G_0(\tau_1,\tau_2)]_{\beta\beta}[G_0(\tau_2,\tau_1)]_{\alpha\alpha} \nonumber \\
&\times \sum_{l}[D_0(\tau_1,\tau_2)]_{l,l}\sum_{i,j}S^i_{nl}S^j_{lm}[{\sigma}^i]_{{\alpha}{\beta}}[{\sigma}^j]_{{\beta}{\alpha}}\:,
\end{align}
where we have evoked the assumption that the electrons are spin degenerate thus omitting the spin index on $G_0(\tau_1,\tau_2)$ and including a factor of 2. We 
now calculate the real-time quantities, such as the lesser (greater) self-energies, by using Langreth's theorem for the time ordering over the defined
contour \cite{Haug}. After including the elastic contribution we obtain
\begin{align}
\label{eq:12}
&[\Pi^{\lessgtr}(t_1,t_2)]^{(2)}_{nm}=-2J^2_\mathrm{sd}\sum_{\alpha,\beta}[G^{\lessgtr}_0(t_1,t_2)]_{\beta\beta}[G^{\gtrless}_0(t_2,t_1)]_{\alpha\alpha} \nonumber \\
&\times \sum_{l}[D^{\lessgtr}_0(t_1,t_2)]_{ll}\nonumber \\
&\times\sum_{i,j}\Big(S^i_{nl}S^j_{lm}[{\sigma}^i]_{{\alpha}{\beta}}[{\sigma}^j]_{{\beta}{\alpha}}+\delta_{ij}\delta_{\alpha\beta}\chi S^i_{nm}[{\sigma}^i]_{{\alpha}{\beta}}\Big)\:.
\end{align}
On computing the Fourier transform we note the two different expressions for the lesser and greater Green's functions are
\begin{align}
\label{eq:14}
&[\Pi^{\lessgtr}(E)]^{(2)}_{nm}=-\frac{J^2_\mathrm{sd}}{\pi}\sum_lP^{\lessgtr}_l\times \nonumber \\
&\sum_{\alpha,\beta}\int\limits_{-\infty}^{+\infty}{d}{\omega}[G_0^{<}(\omega)]_{\beta\beta}[G_0^{>}(\omega\pm(E-\varepsilon_l))]_{\alpha\alpha}\times\nonumber \\ &\sum_{i,j}\Big(S^i_{nl}S^j_{lm}[{\sigma}^i]_{{\alpha}{\beta}}[{\sigma}^j]_{{\beta}{\alpha}}+ \delta_{ij}\delta_{\alpha\beta}\chi S^i_{nm}[{\sigma}^i]_{{\alpha}{\beta}}\Big),
\end{align}
where we have defined $P^<_l=P_l$ and $P^>_l=1-P_l$ and $D_0^{\lessgtr}(t_1,t_2)_{ll}=P^{\lessgtr}_l\text{exp}[{-i\varepsilon_l(t_1-t_2)}/\hbar]$. 
By assuming that the spin system is in thermal contact with a heat bath kept at the temperature $T$, the energy levels $\varepsilon_l$ should be broadened by 
the factor $\beta=k_\mathrm{B}T$. This can be neglected for the ease of the calculation since in general $T<<1$. However we do not disregard the broadening in the 
electronic Green's function due to contact to tip and substrate as this is pivotal to the calculation of the non-equilibrium spin populations.

\end{document}